\documentstyle[12pt]{article}
\begin{document}

\title{{\bf Mechanical Interpretation \\
of Existence Theorems \\
in a Nonlinear Dirichlet Problem}}

\author{Augusto Gonzalez}

\date{Depto. de Fisica\\
      Univ. Nacional de Colombia, Sede Medellin\\
      AA 3840, Medellin, Colombia\\
      and\\
      Instituto de Cibernetica, Matematica y Fisica\\
      Calle E 309, Vedado, Habana 4, Cuba}
\maketitle

\begin{abstract}
The existence of radial solutions of a nonlinear Dirichlet problem in a
ball is translated to the language of Mechanics, i.e. to requirements on
the time of motion of a particle in an external potential and under the
action of a viscosity force. This approach reproduces existing theorems
and, in principle, provides a method for the analysis of the general
case. Examples of new theorems are given, which prove the usefulness of
this qualitative method.
\end{abstract}
\newpage

\section{Introduction}

In the present paper, we consider the following nonlinear Dirichlet
problem

\begin{eqnarray}
\Delta u + f(u) &=& 0 ~~~~ {\rm in} ~~~~ \Omega , \\
u &=& 0 ~~~~ {\rm on} ~~~~ \partial\Omega,
\end{eqnarray}

\noindent
where $f$ is a differentiable function and  $\Omega$ is the ball of
radius $R$ in $\Re^D$. We look for conditions guaranteing the existence
of spherically symmetric solutions to (1-2).

The above mentioned problem has been extensively studied in the last
years (see, for example, [1-5] and references therein). In this paper,
our purpose is to develop a very simple picture, based on Mechanics,
for the analysis of the existence of solutions to (1-2). This
qualitative picture reproduces the existing results and, in principle,
provides a frame for the analysis of the radial solutions to (1 - 2)
in presence of an arbitrary nonlinear function $f$. Examples of new
theorems are given, which show the usefulness of the method.

To our knowledge, the analogy of the radial equation (1) with the
Newtonian law of motion of a particle was first used by Coleman [6]
to obtain the approximate form of the solution connecting false and
true vacua in scalar field theories. This solution enters the
semiclassical expresion for the decay probability of the
false vacuum state. Application of this analogy to the analysis of the
existence of solitary waves in nonlinear one-dimensional media has
proven to be very useful too [7].

The plan of the paper is as follows. In the next Section, the problem
about the existence of solutions to (1-2) is translated to the language
of Mechanics. Two limiting solvable cases, the one-dimensional problem
and the linear equation, are considered and a few general results are
given. Let us stress that the function $f(u)$ is interpreted as the
derivative of a potential, thus the linear equation describes the
motion in a quadratic potential. Section 3 deals with potentials
having a well around $u = 0$. The most interesting examples studied in
this Section are, in our opinion, the potentials with barriers. In
Section 4, we study the motion in a potential with a hill around
$u = 0$. In Section 5, we consider singular
(finite and infinite) potentials. Concluding remarks are given at the
end of the paper.

\section{The Analogy with Mechanics}

We start by considering the spherically symmetric version of Problem
(1-2)

\begin{eqnarray}
{{\rm d}^2 u \over {\rm d}r^2}
 + {D - 1 \over r}{{\rm d}u \over {\rm d}r} + f(u) = 0, \\
 {{\rm d}u \over {\rm d}r} (0) = 0, ~~~~~~ u(R) = 0.
\end{eqnarray}

Written in this form, the analogy with Mechanics is evident. Equation 3
is nothing, but the Newton law for a particle of unit mass moving in a
potential$V$ which is the antiderivative of $f$, $f(u) = {\rm d}V /
{\rm d}u$, and under the action of a viscosity force inversely
proportional to time. The particle should start with zero velocity from
a position $u(0)$ and arrive to $u = 0$ in a time $R$ (Fig. 1a).

We have drawn in Fig. 1b a generic positive solution to (3-4) for a given
$V$. In general, the particle will realize damped oscillations around the
point $u = 0$ (Fig. 2). Let $T_n(u(0))$ be the time the particle spends to
reach the point $u = 0$ $n$ times starting from $u(0)$. Thus, the
existence of a solution to (3-4) may be formulated in the following terms:

``In the potential $V$, there exists an $u(0)$ and a positive integer, $n$,
such that $T_n(u(0)) = R$''

The interesting point is that in many cases we may perform simple
estimates,  based on physical principles, of the dependence $T_n$ vs
$u(0)$ and,
consequently, we may give criteria for the existence of solutions to (3-4).

Let us first study two limiting cases in which equation (3) may be solved
exactly. They will be very useful in the analysis below.

\subsection{The one-dimensional ($D = 1$) problem}

The $D = 1$ case is characterized by the absence of friction. Thus, the
energy $E = (1/2)({\rm d}u/{\rm d}r)^2 + V(u)$ is conserved, ${\rm d}E/
{\rm d}r = 0$, and the dependence $r(u)$ may be expressed in the form of
an integral in each interval where ${\rm d}u/{\rm d}r$ does not change
sign,

\begin{equation}
r - r_a = \int^r_{r_a} ~ {\rm d} t = \{ {\rm sign} ({\rm d}u/{\rm d}r) \}
          \int^u_{u_a} ~ { {\rm d}x \over \sqrt{2( V(u(0)) - V(x))} }.
\end{equation}

In such conditions, the motion of a particle in a well is a periodic
motion characterized by the function $T_1$

\begin{equation}
T_1(u^+(0)) = \int^0_{u^+(0)}~{{\rm d}x \over \sqrt{2(V(u^+(0))-V(x))}}.
\end{equation}

\noindent
(For negative $u(0)$ the integration limits shall be reversed). Note that
$T_n$ may be expressed in terms of $T_1$:

\begin{equation}
T_n (u^+(0)) = \bigg( 2[{n + 1 \over 2}] - 1 \bigg) T_1(u^+(0))
               + 2 [{n \over 2}] T_1(u^-(0)),
\end{equation}

\noindent
where $[q]$ means the integer part of $q$, and $u^-(0)$ is defined from
$V(u^+(0)) = V(u^-(0)) = E$.

For a given potential, the equation $T_n = R$ may be explicitly written
and the existence of solutions to Problem (3 - 4) may be explicitly
investigated.

We are not going to give further details of the analysis in this simple
case and turn out to the higher dimensional ($D > 1$) problem, i.e.
motion with friction. In this situation, there is another exactly
solvable problem: the motion in a quadratic potential.

\subsection{Motion in a quadratic potential (The linear equation)}

We consider a quadratic potential $V(u) = (1/2) \lambda u^2$. The equation
of motion (3) takes the form

\begin{equation}
{{\rm d}^2 u \over {\rm d}r^2} =
 - \lambda u - {D - 1 \over r}{{\rm d}u \over {\rm d}r} .
\end{equation}

The solution of this Eq. with initial condition
${{\rm d}u \over {\rm d}r} (0) = 0$ is expressed as $r^{1 - n/2}
J_{|D/2 -1|}(\sqrt{\lambda} r)$, where $J$ is the Bessel function [8].
It is important to note that the main properties of the solution may be
understood simply from the invariance properties of Eq. (8).

LEMMA: $T_n$ does not depend on $u(0)$ and is proportional to
$\lambda^{-1/2}$.

PROOF: The Eq. is invariant under a change in the scale of $u$, and
also under the transformation

$$r \to C_r r, ~~~\lambda \to C_{\lambda} \lambda,$$

\noindent
where $C_r = C_{\lambda}^{-1/2}$.

According to this Lemma, the function $T_n(u(0))$ takes a fixed value
that depends only on $\lambda$ and $n$. Varying appropiately the parameter
$\lambda$ (the potential), one may fulfil the requirement $T_n = R$. The
corresponding set of parameters , $\{ \lambda_n \}$, define the
eigenvalues of the linear problem.

\subsection{Some useful results}

In this Subsection, we derive a few general results following from the
analogy with Mechanics and classify the potentials to be studied.

In the presence of dissipation, the rate of change of the energy is
written as

\begin{equation}
{\rm d}E/{\rm d} r ~=~ {{\rm d} \over {\rm d}r} \bigg( (1/2)
                     ({\rm d}u/{\rm d}r)^2 + V(u) \bigg) ~=~
                     -{D - 1 \over r} ({\rm d}u/{\rm d}r)^2 ~<~ 0,
\end{equation}

\noindent
i.e. $u(r)$ is damped, as mentioned above. It means that
$E(u(0)) = V(u(0)) > E(0) > V(0) = 0$ (we have supposed that $f$ is
integrable, so that $V(u)$ may be defined as $\int_0^u ~ f(x) {\rm d}x$).
Then, we arrive at the following

THEOREM (A necessary condition): If $u(r)$ is a solution to (3 - 4) and
$f$ is integrable, then $u(0)$ is such that

$$\int_0^{u(0)} ~ f(x) ~{\rm d}x > 0, ~~~ {\rm sign} (f(u(0))) =
{\rm sign}(u(0)).$$

The last condition on the sign of $f(u(0))$ means that the particle shall
be pushed towards the origin at the initial position, $u(0)$. Otherwise,
it will never move to the origing passing through $u(0)$ because of the
energy losses.

More sophisticated versions of this Theorem will be formulated below when
studying potentials with barriers.

A second important result concerns the retardation effect of friction. Let
us suppose that the particle moves from $u_a$ to $u_b$. The time it spends
in this motion may be written as

\begin{equation}
r_b - r_a = \int^{u_b}_{u_a}~{ {\rm d}x \over \sqrt{ ({\rm d}u_a/
            {\rm d}t_a)^2 + 2 V(u_a) - 2 V(x) - 2 (D - 1)
            \int_{r_a}^t ~ {{\rm d}\tau \over \tau} ({\rm d}x /{\rm
            d}\tau)^2} }.
\end{equation}

Of course, this is not a closed expression because the derivative in the
time interval $(r_a, r_b)$ enters the r.h.s. of it. However, it is
evident that

\begin{equation}
r_b - r_a > \int^{u_b}_{u_a}~{ {\rm d}x \over \sqrt{ ({\rm d}u_a/
            {\rm d}t_a)^2 + 2 V(u_a) - 2 V(x) } },
\end{equation}

\noindent
i.e.

LEMMA: The time interval $r_b - r_a$ is greater than the time the
particle spends to move from $u_a$ to $u_b$ without friction.

Finally, let us classify the potentials according to their properties in
the neighborhood of $u = 0$. In the present paper, we will study four
classes of potentials having different behaviours in the vicinity of
this point (Fig. 3):

\begin{description}

\item{a)} The wells are defined as concave potentials around $u = 0$.
\item{b)} The hills are convex around $u = 0$. Of course, at large
$|u|$, $V(u)$ shall be positive (the necessary condition).
\item{c)} and d) correspond to singular potentials.

\end{description}

We will study below each class of potentials separately.

\section{Wells around $u = 0$}

A well is defined as a region with only one local extremum, the minimum
at $u = 0$. In this Section, we study some examples of potentials having
a well around $u = 0$.

\subsection{Potentials, quadratic in $u = 0$ and $|u| \to \infty$}

Let $V(u)$ be a potential such that

\begin{eqnarray}
\left. V \right|_{u \to 0} &\approx& \frac{1}{2} \lambda(0) u^2,
     \nonumber \\
\left. V \right|_{|u| \to \infty} &\approx& \frac{1}{2} \lambda(\infty)
     u^2,
\end{eqnarray}

\noindent
additionally, we will asume that the only zero of $f$ is at $u = 0$.
Then, we have the following

THEOREM: If $\lambda(0) < \lambda_1$ and  $\lambda(\infty) > \lambda_k$,
then Problem (3 - 4) has at least $2 k + 1$ solutions.

This Theorem was obtained in [5]. We will give a detailed proof of it by
means of our method as an illustration. For the incomming Theorems, the
proof will be shortened.

The statement is that the function $T_n$ vs $u(0)$ has the form depicted
in Fig. 4 for $1 \le n \le k$, i.e. for each $T_n$ there are two
solutions.

Indeed, the very small amplitude motion is governed by the $u \to 0$
asymptotics of $V$. $T_n$ depends very smoothly on $u(0)$ in this region
and $T_n \ge T_1 > R$. The latter inequality comes from $\lambda(0) <
\lambda_1$.

On the other hand, the large amplitude motion is governed by the $|u|
\to \infty$ asymptotics and, according to the inequality $\lambda(\infty)
> \lambda_k$, we have $T_n \le T_k < R$. The point to clarify is why
$T_n$ for large $u(0)$ is not affected by the small-$u$ behaviour of $V$.

The answer is that, when $u(0)$ is large, the time the particle spends to
move in the small-$u$ region is negligible. This result comes from the
scale invariance of the quadratic potential as shown in Fig. 5. Shadowed
areas correspond to motion in the region $|u| < u_a$. It is seen that
when $|u(0)| \to \infty$ the time spent in this motion shrinks to zero.
It means that one can deform $V(u)$ at low $|u|$ without changing
significantly $T_n$.

Thus, Problem (3 - 4) has $2 k$ nontrivial solutions plus the trivial
$u = 0$.

\subsection{Potentials with barriers}

In the previous Subsection, we assumed continuity of $T_n$ vs $u(0)$.
However, continuity is broken when $V$ has local extrema, others than
$u = 0$. The point is that, as may be seen from Eq. 5 and the retardation
Lemma of Section 2.3, the time the particle spends to move out of a local
maximum tends to infinity when $u(0)$ approaches the position of the
maximum.

Then, let us first suppose that $f$ has a unique second zero at a point
$a > 0$. The following Theorem may be formulated

THEOREM. If $\lambda(0) > \lambda_k$ and $\left. V \right|_{u(0) \to
-\infty} > V(a)$, then Problem (3 - 4) has at least $k$ solutions with
$0 < u(0) < a$.

To prove it, we draw again the function $T_n(u(0))$, with $1 \le n \le k$
and positive $u(0)$. At low $u(0)$, $T_n \le T_k < R$. When $u(0) \to a$
from below, $T_n \to \infty$. The condition on $V(-\infty)$ guarantees
that the motion is oscillatory around $u(0)$ and the particle does not
escape to $- \infty$.

Note that it is difficult to draw the dependence $T_n$ vs $u(0)$ for
negative $u(0)$ without a knowledge of the potential. The following
Theorem, contained in Ref. [5], states that for asymptotically
quadratic potentials one can say much more.

THEOREM. If $f$ has positive zeroes, the first of which is at $u = a$,
and $\lambda(0)$, $\lambda(\infty) > \lambda_k$, then Problem (3 - 4)
has at least $4 k - 1$ solutions.

We have drawn in Fig. 6 the potential and the functions $T_1$, $T_n$,
$1 < n \le k$. The points $b_+$ and $b_-$ are defined in the monotone
regions. They satisfy $V(a) = V(b_+) = V(b_-)$. Dashed lines means that
the curves are conditionally drawn, while shadowed intervals of $u(0)$
mean physically impossible initial conditions.

The dependence $T_1$ on $u(0)$ when $0 < u(0) < a$ is the same as in the
previous Theorem. For very large positive $u(0)$, $T_1$ is determined by
$\lambda(\infty)$, i.e. $\left. T_1 \right|_{u(0) \to \infty} < T_k < R$.
On the other hand, because of energy losses, if the particle starts from
$b_+$ it will not reach the origin. By continuity, there exists $c_1 >
b_+$ such that the particle arrives at $a$ with zero velocity. This
corresponds to an infinite $T_1$. When $u > c_1$ the particle reaches the
origin and the dependence $T_1(u(0))$ is shown. Note that we can not say
anything about $T_1$ for negative $u(0)$. Thus, the equation $T_1 = R$
will have, at least, two solutions.

Analog reasonings are used in the analysis of $T_n$, $1 < n \le k$. $C_n$
is now defined such that when $u(0) > c_n$ the origin is reached $n$
times. Note that $T_n(c_n) = \infty$ and also that $c_1 = c_2 < c_3
= c_4 < c_5 \ldots$. On the l.h.s. of the origin, we can define the
points $e_n < d < b_-$. $d$ is such that when the particle arrives to $a$
it do so with zero
velocity, while $e_n$ is such that for $u(0) < e_n$, the particle reaches
the origin $n$ times. Note that $e_2 = e_3 > e_4 = e_5 > e_6 \ldots$. In
other words, for each $n$ there are 4 solutions. This proves the Theorem.

Notice that, unlike papers [1 - 5], we are able to indicate forbidden
regions for $u(0)$. This is a generalization of the necessary condition of
Section 2.3.

\subsection{The potentials $V = g |u|^\beta$}

Let us now consider the potentials $V = g |u|^\beta$, with $g > 0$, $\beta
> 1$. We shall first prove that, whatever $\beta$ be, $u(r)$ will have the
form drawn in Fig. 2. After that, we will use scale-invariance properties
of the equation of motion to obtain the dependence $T_n$ vs $u(0)$. Let us
prove the following general

LEMMA. In a potential well, $u(r)$ is an oscillating function of decaying
amplitude.

PROOF. It is evident that the particle will reach the origin whatever the
initial position be. It can not stop in an intermediate point where the
force is not zero. Thus, the question is how long it takes to reach the
origin and what is the final velocity. If this time and the velocity are
finite, we can repeat the argument to conclude that $u(r)$ will have
infinite zeroes.

Let $r_a$ be an intermediate time such that $\left |{{\rm d}u \over
{\rm d}r}(r_a) \right | > 0$. Due to the particular form of the friction,
we can obtain an upper bound for the time to reach the origin starting
from $u(r_a)$, $r_b$, and a lower bound for $\left |{{\rm d}u \over
{\rm d}r} (r_b) \right |$, if we neglect the potential for $r > r_a$ and
solve the problem:

\begin{eqnarray}
{{\rm d}^2 u \over {\rm d}r^2} + {D - 1 \over r}{{\rm d}u \over {\rm d}r}
        = 0 ~~~, \\
u(r_a) = u_a, ~~~ {{\rm d}u \over {\rm d}r}(r_a) = v_a ~~~,
\end{eqnarray}

\noindent
which has the following solution

\begin{eqnarray}
{{\rm d}u \over {\rm d}r}(r) &=& v_a (r_a / r)^{D - 1} ~~~, \\
u(r) &=& u_a + v_a {r_a^{D - 1} \over D - 2} \left \{ r_a^{2 - D}
       - r^{2 - D} \right \} ~~~.
\end{eqnarray}

It means that $u = 0$ will be reached in a finite $r_b$ with a finite
velocity and $u(r)$ will have infinite zeroes.

Thus, let us now turn out to the dependence $T_n$ vs $u(0)$ in the
potentials $V = g |u|^\beta$. The equation of motion takes the form

\begin{equation}
{{\rm d}^2 u \over {\rm d}r^2} = - g \beta ({\rm sign}~~ u)
    |u|^{\beta - 1} - {D - 1 \over r}{{\rm d}u \over {\rm d}r}  ~~~.
\end{equation}

The properties of $T_n$ following from the scale invariance of the
equation are given in the next Lemma:

LEMMA. For fixed $g$, $T_n \sim |u(0)|^{1 - \beta/2}$, while for fixed
$u(0)$, $T_n \sim g^{- 1/2}$.

Thus, for every $n$, the equation $T_n = R$ will have two solutions, and
we arrive to the following

THEOREM. Problem (3 - 4) with $V = g |u|^\beta$, $g > 0$, $\beta > 1$ has
infinite solutions [3].

One can now combine these with previous results. In quality of example,
let us formulate the following

THEOREM. Let $\lambda(0) < \lambda_k$ and $\left. V \right |_{|u| \to
\infty} \sim |u|^\beta$, with $\beta > 2$, then solutions to Problem
(3 - 4) with any $n \ge k$ zeroes exist.

The curve $T_n$ vs $u(0)$ for $n \ge k$ may be easily drawn in this case.
Note that the dependence of $\left. T_n \right |_{|u(0)| \to \infty}$ on
the low-$u(0)$ properties of $V$ is, for $\beta > 2$, weaker than in the
quadratic potential.

\section{Hills around $u = 0$}

We now study the motion in a potential like that one shown in Fig. 3 b.
For simplicity, we assume that $V$ is quadratic near zero ($\lambda(0)<0$)
and also quadratic at large values of $u$. No additional zeroes of $f$
exist. Then one can formulate the following

THEOREM. If $\lambda(0) < 0$ and $\lambda(\infty) > \lambda_k$, then
Problem (3 - 4) has $2 k + 1$ solutions.

We have drawn in Fig. 7 the curve $T_n$ vs $u(0)$ for $1 \le n \le k$.
The large-$u(0)$ behaviour of it is evident. The points $b_+$ and $b_-$
are the zeroes of $V$. The points $c_n$ and $e_n$ are defined as in the
previous Section, i.e. starting from the right of $c_n$ (the left of
$e_n$) the particle may reach the origin $n$ times. Note that would it
start from $c_n$ ($e_n$), then it would arrive to $u = 0$ with zero
velocity, i.e. $T_n(c_n) = T_n(e_n) = \infty$. Note also that $b_+ < c_1
< c_2 \cdots$, $b_- > e_1 > e_2 > \cdots$. Thus, for each $n$ there are
two solutions and the Theorem is proved.

Other potentials could be analysed, but we think that to show the
advantages of the method the given example is enough.

\section{Singular Potentials}

The main property of the singular potentials, Figs. 3 c) and d), is that
the force, $- {\rm d}V / {\rm d}u$, at $u = 0$ is ill-defined. So, the
motion an interval of time after the particle reaches the origin is not
well defined, and we can only analyse the existence of positive solutions
to (3 - 4).

An example of a potential like 3 c) is $V = g |u|^\beta$, with $g > 0$
and $0 < \beta < 1$. Let us stress that the upper bound for $r_b$ and the
dependence $T_1 \sim |u(0)|^{1 - \beta/2}$, obtained in the Lemmas of
Section 3.3, are valid, so that the equation $T_1 = R$ has always a
solution in this case.

The same analysis holds for the potential $V = - g |u|^{- \beta}$, with
$g, \beta > 0$. This is a potential of the form 3 d). Scale invariance in
this case leads to $T_1 \sim |u(0)|^{1 + \beta/2}$, so that the equation
$T_1 = R$ will always have a solution also.

We can now combine possibilities to obtain interesting situations. Let,
for example, the potential $V$ be quadratic at the origin with $\lambda(0)
>0$, while at long distances $V \sim V_0 - g |u|^{- \beta}$. No zeroes of
$f$ exist, except the trivial at $u = 0$. Then, we obtain the following

THEOREM. If $\lambda(0) > \lambda_k$, then Problem (3 - 4) has at least
$2 k + 1$ solutions.

The proof is trivial.

\section{Concluding Remarks}

In the present paper, we used the analogy of Eq. (3) with the second
Newton's law in order to obtain existence theorems in Problem (3 - 4).
Appart from reproducing existing results, we give new examples of
potentials (of $f$) in which it is relatively easy to analyse the
existence of solutions.

We think that the given examples show that the method is general enough to
provide a first insight to the problem for any reasonable function $f$.
After that, we may go further on in two ways: i) Use more rigurous methods
to complete the proof and/or ii) Obtain numerical solutions to the
equation.
\vspace{1cm}

{\bf Acknowledgements}

\noindent
The author is grateful to J. Cossio for a presentation of the results of
[5], which motivated the present work, and to the participants of the
Theoretical Physics Seminar at the Universidad Nacional de Colombia, Sede
Medellin, for discussions. The support by CINDEC to the Seminar is
gratefully acknowledged.

\newpage

Figure Captions
\vspace {.5 cm}

Fig. 1

a) The analogy with Mechanics.

b) A positive solution to (3 - 4) corresponding to the situation depicted
in a).
\vspace{.5 cm}

Fig. 2. A generic damped oscillating function $u(r)$ describing the motion
of a particle in $V$.
\vspace{.5 cm}

Fig. 3. Different possibilities for the neighborhood of $u = 0$.

a) Well

b) Hill

c) Finite, but singular

d) Infinite, singular potential
\vspace{.5 cm}

Fig. 4. Dependence $T_n$ vs $u(0)$ for the potential considered in
Section 3.1.
\vspace{.5 cm}

Fig. 5. A consequence of the scale invariance of the quadratic potential.
The shadowed areas correspond to motion in the region $|u| < u_a$. When
$|u(0)| \to \infty$, the time spent in this motion shrinks to zero.
\vspace{.5 cm}

Fig. 6. A potential with barriers and the corresponding $T_1(u(0))$,
$T_n(u(0))$, $1 < n \le k$.
\vspace{.5 cm}

Fig. 7. The curves $T_n$ vs $u(0)$, $1 \le n \le k$, for the potential of
Section 4. Notations are the same as in Fig. 6.

\end{document}